# Gravitational lensing stereoscopy

Ira Rai[a], Vandana Vinayak[b,*], Richard Gordon[c,**]

[a] Indian Institute of Technology, Department of Aerospace Engineering, Madras, India, 60003
[b] Dr. Harisingh Gour Central University, School of Applied Sciences, Sagar, Madhya Pradesh, India, 470003
[c] Retired, University of Manitoba, Department of Radiology, Alonsa, Manitoba, Canada R0H 0A0

**Abstract**. A galaxy cluster, such as RX J2129, sometimes produces two or more gravitationally lensed images of more distant galaxies. We attempt to regard pairs of these images as stereo pairs. While not successful due to the small disparity angles involved, we suggest that with the $10^{11}$ light amplification anticipated from the Solar Gravitational Lens (SGL), individual stars of the distant galaxy might be resolved, resulting in 3D images.

**Keywords**: Gravitational lensing, Stereoscopy, Galaxy cluster, RX J2129, James Webb Space Telescope (JWST)

**Corresponding author\***: E-mail: vvinayak@dhsgsu.edu.in (Vandana Vinayak) and DickGordonCan@protonmail.com (Prof. Richard Gordon)

1. ## Introduction

Gravitational lensing has allowed us to see galaxies behind clusters of galaxies that provide gravitational lensing with a large distance between lenses. A remarkable image has been obtained with the James Webb Space Telescope (JWST)(Kelly, Chen, Alfred et al., 2022; Lea, 2023), showing three images lensed by galactic cluster RX J2129 of a galaxy containing Type Ia supernova AT 2022riv (cf.(Morgan, Chartas, Malm et al., 2001)). Due to their different path lengths, two of the three images are 320 days and 1000 days after the first. This allowed the time course of the lensed supernova to be ascertained. But it also allows stereoscopy, as little displacement of stars in the distant galaxy is expected over these astronomically short periods. Thus, the prospect presents itself of making stereo pairs from the three images of a distant galaxy. As the multiple gravitational lenses will generally be distinct, methods for compensating for distortions between the images (deconvoluting them (Chantry and Magain, 2007; Melchior, 2008; Rogers and Fiege, 2012; Turyshev and Toth, 2022) will be necessary. "Distortions are removable in principle if we know the mass distribution of the lens. But that information is not typically available, precise correlation between such images may help recover more information about the lens[es]" (Viktor T. Toth, personal communication, 2023). Simple dewarping (Safjan, 2022) might introduce stereopsis artifacts. As per our knowledge, the stereo pairs reported for JWST in this paper have not been reported till date.

We have previously shown how a translucent object may be reconstructed via a very few projection angles and observed in 3D via stereoscopy(Jaman, Gordon and Rangayyan, 1985). This should be generalizable to self-luminant objects.

## 2. Methods

The 3 inserts from the JWST image (Kelly, 2023) (Figure1) were cropped.

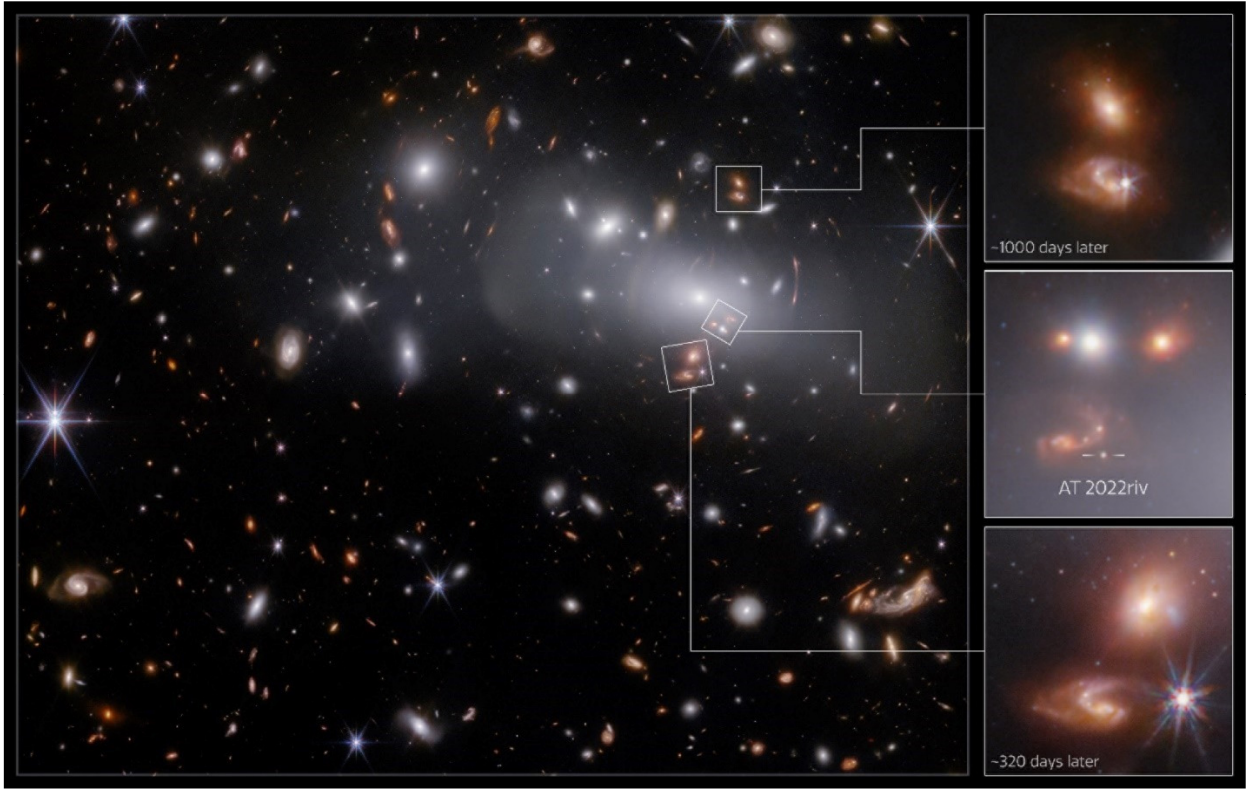

**Figure 1.** James Webb Space Telescope (JWST),(Kelly, 2023).

Two galaxies behind galaxy cluster RX J2129 are seen in three gravitationally lensed images by the Webb Telescope(European Space Agency, 2023). While the emphasis was on the temporal changes in a supernova in the distant galaxy, ours is that these represent three images from three different viewpoints, allowing an attempt at 3D reconstruction of the farthest galaxy. RX J2129 is about 3.2 billion light years from Earth. The original image posted is 5707 × 3621pixels (118.3 MB) subtending 2.25 x 1.80 arcminutes, taken at 1.15µm, 1.5µm, 2.0 µm, 2.77µm, 3.56µm and 4.44µm in the infrared(Kelly, 2023). The magnification and distance of the distant galaxy have not yet been reported. The galaxy cluster is reported at $3.2 \times 10^9$ light years from Earth(Kelly, 2023).

We further processed the bottom distant galaxy, showing the most structure. The higher resolution insert images from JWST were cropped and resized to 195×195 pixels (Figure ), using the geometrical centroids for the pair of galaxies. Further cropping of these images excluded stars presumed to be in the foreground. The Euclidean norm for each pair was calculated versus relative angle of rotation about the centroids. The angle giving minimum norm was used to form stereo pairs. Note that while the minimum norm could be a measure of disparity, it also measures gravitational lens distortions, for which we did not correct. Coding was in Python using Libraries PIL, NUMPY, MATPLOTLIB, SCIPY, and CV2.

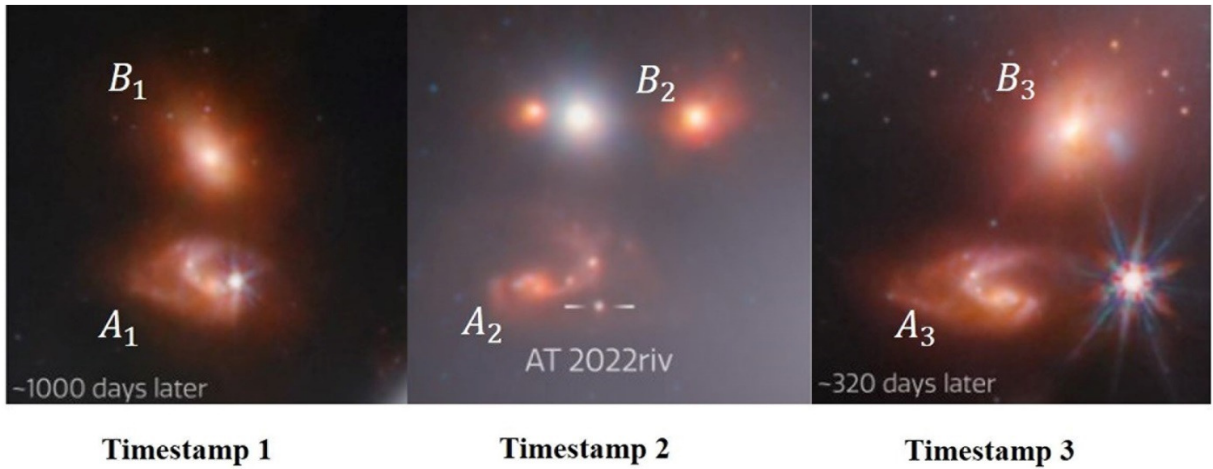

**Figure 2.** The 3 inserts of JWST resized to 195×195 pixels, marking 2 galaxies (A and B) at 3 different timestamps, named as $A_1$, $A_2$, $A_3$, $B_1$, $B_2$, $B_3$.

**Note*** The middle image $A_2$ appears rotated by its gravitational lens compared to the other two. Image $A_1$ shows the supernova. Image $A_3$ shows the same galaxy at the bottom through a gravitational lens different from the other two.

## 3. Results

The minimum disparity of 9 squared pixels was found between images **$A_1$** and **$A_2$** (Figure ), using an algorithm following MSE (mean squared error method). The values were 53 for the **$A_1\,A_3$** pair and 97 for the **$A_2\,A_3$** pair. The images for galaxies in above inserts was cropped as galaxy $A_1$, $A_2$, $A_3$, $B_1$, $B_2$ and $B_3$. Each cropped image was resized to the size of image with higher pixel dimension.

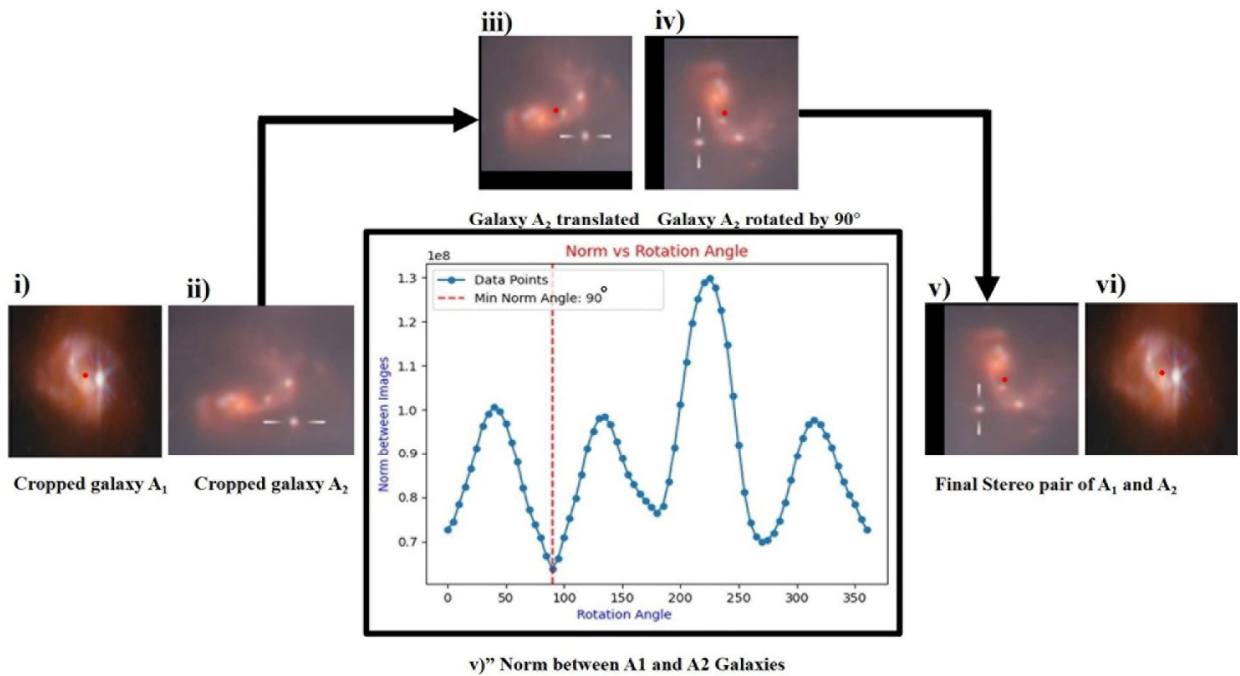

**Figure 3.** Process showing modification of Galaxy $A_2$ such that galaxies $A_1$ and $A_2$ have minimum norm (denoted by red vertical line in the graph) and act as stereo pairs.

In above figure 3, the process of making a stereo pair between galaxy $A_1$ and $A_2$ is represented. The figure 3 i) and ii) shows the cropped image of galaxy $A_1$ and $A_2$ respectively; figure 3 iii) and ii) is translated to have centroid at same coordinates as that in figure 3i). Further figure 3 iv) shows where figure 3 iii) of galaxy $A_2$ is rotated by 90 ° to obtain minimum norm between figure 3 i) and iii). In figure 3 v) shows graph (figure 3v)'') of norm between images A1 and A2 by a plot between norm vs rotation angle, for every 5 ° of rotation. The graph (figure 3v)'') shows that the minimum norm can be observed as 90 °. Further figure 3vi) represents the stereo pair for galaxy $A_1$ and $A_2$.

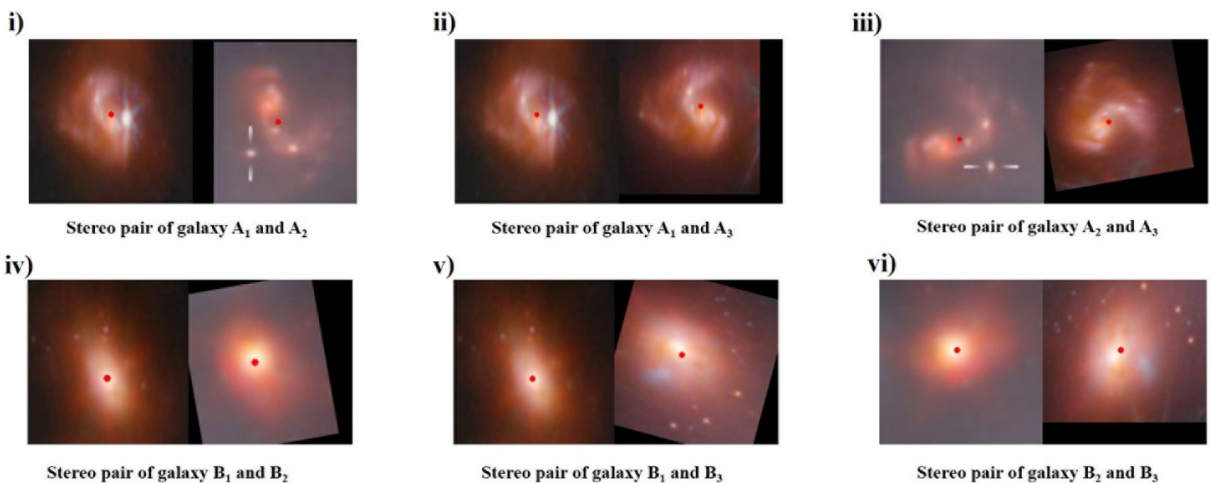

**Figure 4**. Possible stereo pairs for galaxy A as shown in i), ii) and iii) and for Galaxy B as shown in iv), v) and vi).

This is our best approximation to a stereo pair. Structural differences may be due more to gravitational lens aberrations than stereo disparities. Figure 5 further represents stereo images of constellation Tucanae 47, a Galaxy, Geometry sketch of room and images captured of Indian Elephants and their stereo images made using software application (i3Dsterioid).

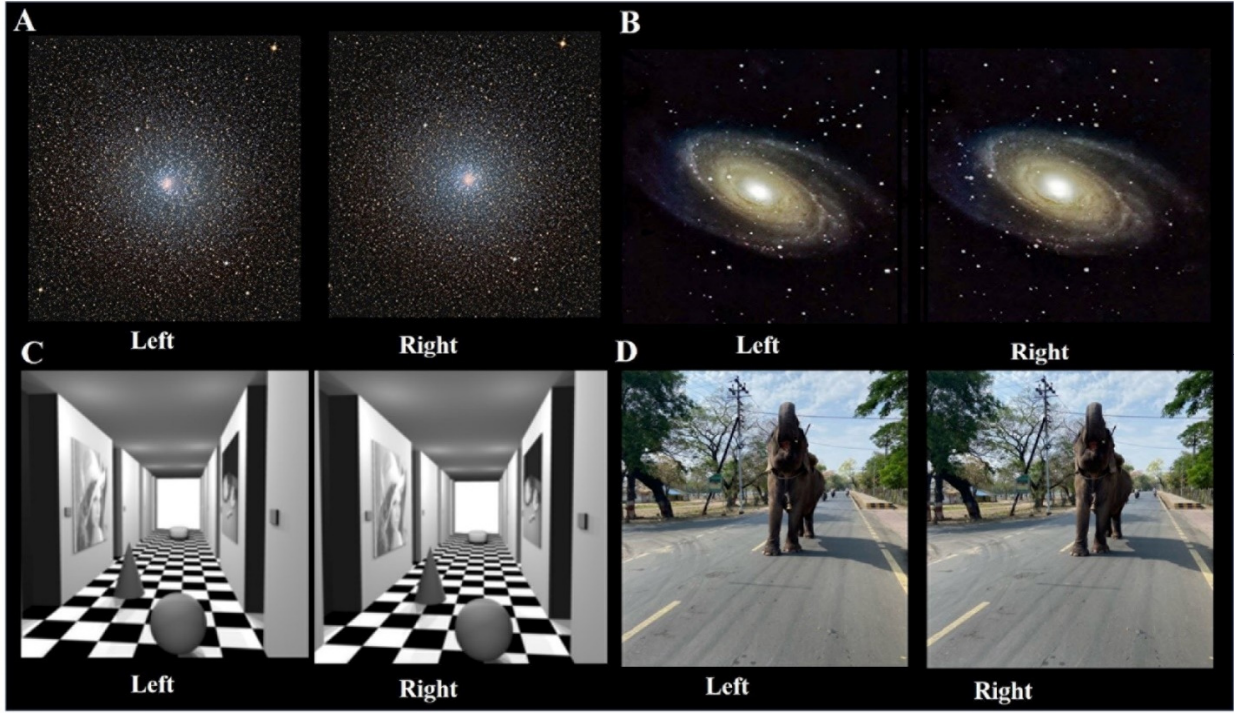

**Figure 5.** Stereo images of **A)** constellation Tucanae 47 **B)** a Galaxy **C)** Geometry sketch of room and **D)** images of Elephants and their stereo images made using software application.

## 4. Discussion

Perfect alignment without stereo disparity would have a minimum norm of 0°. We leave the attempt of correcting for gravitational lens aberrations to others, perhaps with super-resolution(Wang, Li and Kang, 2022). A successful attempt might be made with the proposed SGL(Turyshev and Toth, 2020; Turyshev and Toth, 2022; Helvajian, Rosenthal, Poklemba et al., 2023). While this telescope is designed with exoplanet imaging in mind, its anticipated $10^{11}$ magnification/light amplification applies to the whole sky. With the multiple gravitational lenses available from galaxy clusters, which provide another factor of 10 to 100 in magnification/light amplification(Pascale, Frye, Diego et al., 2022), it is a simple calculation to show that individual stars averaging 2 light years apart will sometimes be seen in imaged galaxies. The combination of galaxy cluster imaging with the SGL is analogous to a compound stereo microscope, albeit with magnifications of the objective and eyepiece reversed.

The longest baseline available to us for now is our rotation around the Milky Way (Leong, 2002) (a period whose average estimate is 188 million years. In the long run, if the Universe keeps expanding, distant galaxies will no longer be visible:

> "We better observe the Universe in the next tens of billions of years and document our findings for the benefit of future scientists who will not be able to do so"(Loeb, 2015).

Thus, we could assemble images soon that up to 94 million years from now might be of use as one of the two images needed for stereometry. In the interim, every gravitational lens will sweep an arc and may occasionally bring new distant objects into view, and so may be worthy of long-term study.

## 5. Conclusions

Hence, we propose that 3D imaging of distant galaxies will be available, if not now, with the SGL (Solar Gravitational Lens). This appears to be a new idea. There are many more cases of multiple gravitational images of the same object that might be objects for stereology (58 references available on request). It proposes a compound telescope consisting of pairs of images obtained from galaxy clusters viewed with the SGL, bigger than any other telescope yet conceived.

**Acknowledgments**

We would like to thank Patrick Kelly (Kelly, Broadhurst, Chen et al., 2022) for the image that inspired this idea, and for suggesting that it has not yet been tried, and William E. Harris and Viktor T. Toth for their assistance, critical comments and probably comment that this idea is worth exploring further w.r.t such as the size of the projected image, the resolution needed to achieve for the stereoscopic images, the anticipated brightness, the requisite integration times to remove, e.g., corona noise, analysis on how to resolve the degeneracy between the multipole moments of the lens and the stereoscopic differences between its projections. We look forward for working on the critical comments of Viktor Toth in our future manuscript and work. IR is thankful to IIT Madras, India for the open environment of study and learning and Prof A.K Dhawan for gifting a Stereo viewer to analyse the stereo images.

**Disclosures**
The authors have no relevant financial interests in the manuscript and no other potential conflicts of interest to disclose.

Data sharing is not applicable to this article, as no new data were created or analyzed.

## References

Chantry, V., & Magain, P., 2007, Astron. Astrophys. 470(2): 467.

European Space Agency, 2023, Seeing triple (annotated).

Helvajian, H., Rosenthal, A., Poklemba, J., et al., 2023, Journal of Spacecraft and Rockets 60(3): 829.

Jaman, K.A., Gordon, R., & Rangayyan, R.M., 1985, Computer Vision, Graphics, Image Processing 30(3): 345.

Kelly, P., 2023, Seeing Triple (Annotated).

Kelly, P., Broadhurst, T.J., Chen, W., et al., 2022: #17253.

Kelly, P.L., Chen, W., Alfred, A., et al., 2022, arXiv preprint arXiv:2211.02670.

Lea, R., 2023, Physics World 36(3): 41.

Leong, S., 2002, Period of the Sun's Orbit around the Galaxy (Cosmic Year).

Loeb, A., 2015, From the First Star to Milkomeda. (Amazon Digital Services).

Melchior, P., 2008, Unveiling shapes: Shapelets for galaxy morphology and gravitational lensing studies. in International Conference on Classification and Discovery in Large Astronomical Surveys, (Amer Inst Physics, Ringberg Castle, Germany, 156).

Morgan, N.D., Chartas, G., Malm, M., et al., 2001, Astrophysical Journal 555(1): 1.

Pascale, M., Frye, B.L., Diego, J., et al., 2022, Astrophys. J. Lett. 938(1).

Rogers, A., & Fiege, J.D., 2012, Astrophysical Journal 759(1).

Safjan, K., 2022, 15 tools for document Deskewing and Dewarping.

Turyshev, S.G., & Toth, V.T., 2020, Phys. Rev. D 101(4).

Turyshev, S.G., & Toth, V.T., 2022, Phys. Rev. D 105(2): 2.

Turyshev, S.G., & Toth, V.T., 2022, Mon. Not. Roy. Astron. Soc. 515(4): 6122.

Wang, L., Li, G.L., & Kang, X., 2022, Mon. Not. Roy. Astron. Soc. 517(1): 787.

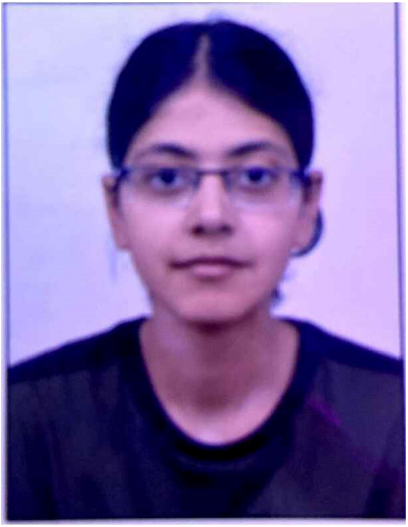

**Ira Rai**

Ira Rai is a student at Indian Institute of Technology, Madras, India in the Department of Aerospace Engineering. Her research interest is, space exploration, neural spiking, machine learning and artificial intelligence to solve unsolved mysteries in space and other fields. She has published as lead author in *Sustainable Energy Technologies and Assessments*. She has a Trademark/Patent on SPACE$^M$ (Life on Mars). She is a space enthusiast (https://www.abhyudayiitm.com/team), selected for Spaceport America competition 2023 (10-24$^{th}$ June'2023 at Las Cruces, US), and worked on the flight computer subsystem for the CANSAT, India competition, which was selected and ranked 16 among 150 countries which launched their sounding rocket at Las Cruces, United States of America.

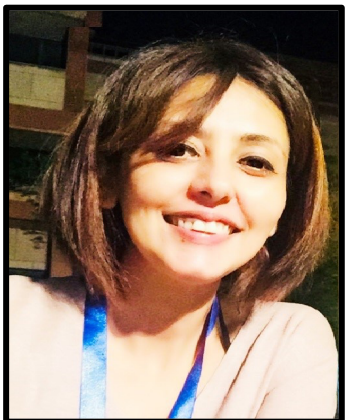

**Dr Vandana Vinayak**

Dr. Vandana Vinayak (www.vandanavinayak.com) is an Assistant Professor at School of Applied Sciences, Dr. Harisingh Gour Central University Sagar, India. Her latest research is towards bubble farming in a Martian environment for terraforming Mars planet. She has published more than 65 peer reviewed papers, has about 15 awards and more than 10 national and international research projects.

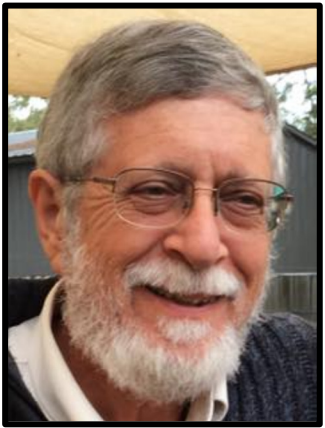
**Prof. Richard Gordon**
Richard Gordon is a theoretical biologist, series editor for Astrobiology Perspectives on Life in the Universe books (Wiley-Scrivener), writing a book on the role of Archaea in the origin of life. He developed the ART computed tomography algorithm, used in solar corona, atmospheres, and seismic 3D reconstruction. He was Professor of Radiology and ECE, University of Manitoba.